\def\ee#1{\ifmmode {} \times 10^{#1} \else ${} \times 10^{#1}$\fi}
\def\msun{{\ifmmode M_\odot \else {$M_{\odot}$}\fi}}
\def\Dnu{{\ifmmode \Delta\nu \else {$\Delta\nu$}\fi}}
\def\DtB{{\ifmmode {\Delta t}_B \else {${\Delta t}_B$}\fi}}
\def\nuB{{\nu_{B}}}

\def\nuorb{{\nu_{\rm orb}}}
\def\nuk{\nu_{\rm K}}
\def\nus {{\ifmmode \nu_{\rm s} \else {$ \nu_{\rm s}$}\fi}}
\def\rsp{R_{\rm sp}}
\def\chisqdof{{\chi^2/{\rm dof}}}
\def\vcl{v_{\rm cl}}
\documentstyle[emulateapj,times]{article}

\begin{document}
\tolerance=10000

\submitted{Submitted July 29, 2000}

\lefthead{LAMB AND MILLER}
\righthead{CHANGING KILOHERTZ QPO FREQUENCY SEPARATION}

\title{Changing Frequency Separation of Kilohertz Quasi-Periodic Oscillations in the Sonic-Point Beat-Frequency Model}

\author{Frederick K. Lamb\altaffilmark{1,2,3} and
        M. Coleman Miller\altaffilmark{4}}

\begin{abstract}
Previous work on the sonic-point beat-frequency (SPBF) model of the kilohertz quasi-periodic oscillations (QPOs) observed in the X-ray flux from neutron stars in low-mass binary systems has shown that it naturally explains many properties of these QPOs. These include the existence of just two principal QPOs in a given source, the commensurability of the frequency separation \Dnu\ of the two kilohertz QPOs and the spin frequency \nus\ inferred from burst oscillations, and the high frequencies, coherence, and amplitudes of these QPOs. Here we show that the SPBF model predicts that \Dnu\ is less than but close to \nus, consistent with the observed differences between \Dnu\ and \nus. It also explains naturally the decrease in \Dnu\ with increasing QPO frequency seen in some sources and the plateau in the QPO frequency--X-ray flux observed in \hbox{4U~1820$-$30}. The model fits well the QPO frequency behavior observed in \hbox{Sco~X-1}, \hbox{4U~1608$-$52}, \hbox{4U~1728$-$34}, and \hbox{4U~1820$-$30} ($\chisqdof = 0.4$--2.1), giving masses ranging from 1.59 to $2.0\,\msun$ and spin rates ranging from 279 to 364$\,$Hz. In the SPBF model, the kilohertz QPOs are effects of strong-field gravity. Thus, if the model is validated, the kilohertz QPOs can be used not only to determine the properties of neutron stars but also to explore quantitatively general relativistic effects in the strong-field regime.
\end{abstract}


\altaffiltext{1}{Center for Theoretical Astrophysics, University of Illinois at Urbana-Champaign, 1110 W. Green Street, Urbana, IL 61801-3080.} 
\altaffiltext{2}{Department of Physics, University of Illinois at Urbana-Champaign.} 
\altaffiltext{3}{Department of Astronomy, University of Illinois at Urbana-Champaign.}
\altaffiltext{4}{Department of Astronomy, University of Maryland at College Park, College Park, MD 20742-2421.}

\section{INTRODUCTION}

Kilohertz quasi-periodic brightness oscillations (QPOs) have been discovered in the accretion-powered emission of more than twenty neutron star low-mass X-ray binaries using the {\it Rossi X-ray Timing Explorer\/} ({\it RXTE}) (see van der Klis 2000 for a review). These kilohertz QPOs have frequencies ranging from $\sim$500~Hz up to $\sim$1200~Hz, amplitudes $\sim$1\%--15\% rms in the 2--60~keV band of the  {\it RXTE} Proportional Counter Array, and quality factors $Q \equiv \nu/{\rm FWHM}$ as large as $\sim100$--200. They often appear as two simultaneous peaks in power spectra of the X-ray brightness, at frequencies $\nu_1$ and $\nu_2$ ($>\nu_1$) that move together and can shift upward and downward in frequency by up to a factor of two within a few hundred seconds, apparently as a consequence of changes in the accretion rate. The frequencies of the kilohertz QPOs lie within the range expected for orbital motion near neutron stars and are almost certainly strong-field general relativistic phenomenon (see Lamb, Miller, \& Psaltis 1998; Miller 2000; van der Klis 2000).

In six of the kilohertz QPO sources, strong (amplitudes up to 50\% rms) brightness oscillations with frequencies of hundreds of Hertz have been observed during thermonuclear X-ray bursts. There is very strong evidence that these oscillations are produced by rotation of brighter regions of the stellar surface with the spin of the star and that the most prominent oscillation frequency is the stellar spin frequency $\nus$ or its first overtone (see, e.g., Strohmayer \& Markwardt 1999; Miller 1999). The coherent oscillations observed in burst tails, after hot matter has spread around the star, and the antipodal brighter regions observed in \hbox{4U~1636$-$53} are strong evidence that these stars have dynamically important surface magnetic fields (Miller 1999), as expected on evolutionary grounds (Miller, Lamb, \& Psaltis 1998, hereafter MLP98; Lorimer 2000). In the four sources where burst oscillations and two simultaneous kilohertz QPOs have both been detected with high confidence, the difference $\Dnu \equiv \nu_2-\nu_1$ between the frequencies of the kilohertz QPOs agrees with the fundamental burst oscillation frequency to within 0.7\% to 15\% (see van der Klis 2000).

The properties of the kilohertz QPOs strongly indicate that the upper kilohertz QPO is an orbital frequency and that the lower is generated by the beat of this frequency with the star's spin (Lamb et al.\ 1998; MLP98; Miller, Lamb, \& Cook 1998): (1)~The frequencies of the kilohertz QPOs are in the range expected for orbital motion near neutron stars, consistent with one being an orbital frequency. (2)~The relatively small variation of \Dnu\ in a given source and the approximate commensurability of \Dnu\ with the stellar spin frequency inferred from burst oscillations indicates that the spin of the star is generating the frequency difference. (3)~The observation of at most two strong kilohertz QPOs in any given source (van der Klis 2000; M\'endez \& van der Klis 2000), rather than three, indicates that only a single sideband is being generated, rather than the two sidebands generated by other mechanisms. (4)~In order to beat with the stellar spin, the other motion must be rotation about the center of the neutron star, indicating that the other motion is orbital motion. (5)~Accretion of angular momentum ensures that the star is spinning in the prograde direction, implying that the frequency $\nu_2$ of the upper kilohertz QPO is close to the orbital frequency $\nuorb$ and that the frequency $\nu_1$ of the lower kilohertz QPO is close to the beat frequency $\nuB \equiv \nuorb - \nus$.

This and other evidence motivated the development of the sonic-point beat-frequency (SPBF) model (MLP98). In this model $\nu_2$ is close to the orbital frequency $\nuorb$ at the sonic radius, where the flow in the disk changes from nearly circular to rapidly inspiraling, and $\nu_1$ is comparable to $\nuB$. The transition to hypersonic radial inflow is an effect of strong-field gravity. In the SPBF model, the low-frequency QPOs seen in the Z and atoll sources are produced by the magnetospheric beat-frequency mechanism (Alpar \& Shaham 1985; Lamb et al.\ 1985). The model explains naturally many features of the kilohertz QPOs in addition to their frequencies and frequency separation and the existence of only two principal QPOs in a given source (MLP98). In the SPBF model, the kilohertz QPOs are a strong-field general relativistic effect and are therefore sensitive probes of the properties of the spacetime near the neutron star, including whether there is an innermost stable circular orbit (ISCO) and the gravitomagnetic torque created by the spin of the neutron star.

Careful analyses have shown that in some kilohertz QPO sources $\Dnu$ decreases systematically by 30--100~Hz, depending on the source, as $\nu_2$ increases by a much larger amount (\hbox{Sco~X-1}: van der Klis et al.\ 1997; \hbox{4U~1608$-$52}: M\'endez et al.\ 1998; \hbox{4U~1735$-$44}: Ford et al.\ 1998; \hbox{4U~1728$-$34}: M\'endez \& van der Klis 1999; see also Psaltis et al.\ 1998) and that the separation frequency in \hbox{4U~1636$-$536} is slightly but significantly smaller than the spin frequency inferred from its burst oscillations (M\'endez, van der Klis, \& van Paradijs 1998).

Here we show that in the SPBF model, the inward radial velocity of the accretion flow increases $\nu_1$ and decreases $\nu_2$, making $\Dnu$ smaller than $\nus$. An increase in $\nu_1$ of $\lesssim 5$\% is sufficient to account for the largest observed differences between $\Dnu$ and $\nuB$ and decreases of $\Dnu$ with increasing $\nu_2$. The inward radial velocity of the accretion flow was included in the gas dynamical and radiation transport calculations reported by MLP98, but the effect of this motion on the frequencies of the kilohertz QPOs was neglected. In \S~2 we discuss the generation of the kilohertz QPOs in the SPBF model, derive general expressions for their frequencies, taking into account the inward motion of the accretion flow, summarize the results of numerical simulations, and introduce the model of the gas dynamics in the SPBF model that we use here to calculate the frequencies of the kilohertz QPOs. In \S~3 we show that despite its simplicity, this  model fits accurately the frequency behavior observed in \hbox{Sco~X-1}, \hbox{4U~1608$-$52}, \hbox{4U~1728$-$34}, and \hbox{4U~1820$-$30} and gives reasonable masses and spin rates for these neutron stars.

\section{Kilohertz QPOs in the Sonic-Point Model}

{\em Generation of kilohertz QPOs}.---In the SPBF model, the kilohertz QPOs are produced by clumping of gas in the accretion flow near the star, probably caused by interaction of the accreting gas with the weak stellar magnetic field (see Ghosh \& Lamb 1979; Aly \& Kuipers 1990). Only clumps near the sonic radius $\rsp$, where the inward radial velocity of gas leaving the clumps becomes supersonic, contribute to the oscillations (see MLP98). At low accretion rates, radiation drag removes angular momentum from the gas in the clumps, causing it to spiral inward to the stellar surface. At high enough accretion rates, the optical depth from the stellar surface is large and radiation drag is unimportant. If, however, the star is smaller than the ISCO, gas will fall inward supersonically from clumps near the ISCO, where general relativity causes gas to spiral inward even if it does not lose any angular momentum. In either case, the supersonic inflow is a strong-field general relativistic effect.

Where gas falling inward from clumps near $\rsp$ impacts the stellar surface, it produces bright footprints that move around the star with a frequency nearly equal to the orbital frequency $\nuorb$ of the clumps at $\rsp$, causing the X-ray flux seen by distant observers to oscillate as the footprint moves into and out of view. (As discussed by MLP98, $\nuorb$ is not exactly equal to the Keplerian frequency $\nuk$, because the radiation force has a radial component and the clumps are drifting slowly inward.) {\em This oscillation is the upper kilohertz QPO; its frequency $\nu_2$ is determined by the rate at which the azimuthal phase of the footprint advances}. 

The weak magnetic field of the star funnels extra gas toward the magnetic poles, producing a broad beam of extra radiation that rotates with the star. Clumps orbiting near $\rsp$ move through this beam with frequency $\nuB \equiv \nuorb-\nus$. When a clump is illuminated by the beam, gas in it loses angular momentum to the radiation at a faster rate, so more gas falls inward, causing the luminosity of the footprint to oscillate with a frequency close to $\nuB$. {\em This oscillation is the lower kilohertz QPO; its frequency $\nu_1$ is determined by the frequency at which the mass flux onto a footprint oscillates}.

The flux in both oscillations is generated primarily at the stellar surface, where most of the gravitational energy is released, and hence both oscillations can have relatively high amplitudes.

{\em Kilohertz QPO frequencies and frequency separation}.---The kilohertz QPO frequencies predicted by the SPBF model can be calculated as follows (we use Boyer-Lindquist coordinates for easy reference to infinity). The light travel time from the stellar surface to the sonic radius is very small and hence the time between two successive maxima of the mass inflow rate from a given clump is very close to the beat period $\DtB \equiv 1/(\nuorb - \nus)$. Let $t_1$ and $t_2 = t_1 + \DtB$ be the times of successive maxima, let $r_1$ and $r_2$ and $2\pi\phi_1$ and $2\pi\phi_2$ be the radial and azimuthal coordinates of the clump at $t_1$ and $t_2$, let ${\Delta t}_1$ and ${\Delta t}_2$ be the times required for the gas stripped from the clump at $t_1$ and $t_2$ to reach the surface of the star, and let $2\pi{\Delta\phi}_1$ and $2\pi{\Delta\phi}_2$ be the changes in the azimuthal phase of the gas released at $t_1$ and $t_2$ as it falls from the clump to the stellar surface. Then the frequencies of the lower and upper kilohertz QPOs are
\begin{equation}
\label{nu1}
\nu_1={1\over{t_2-t_1+(\Delta t_2-\Delta t_1)}}
\end{equation}
and
\begin{equation}
\label{nu2}
\nu_2={\phi_2-\phi_1+(\Delta\phi_2-\Delta\phi_1)\over{
t_2-t_1+(\Delta t_2-\Delta t_1)}} \;.
\end{equation}

Suppose first that the inward motion of the clumps near $\rsp$ can be neglected. This is the approximation used in MLP98 to compute the frequencies of the kilohertz QPOs. Then the time required for gas to spiral inward from a given clump to the stellar surface and the change in the azimuthal phase of the gas during its inspiral do not depend on when the gas separated from the clump, i.e., ${\Delta t}_2 = {\Delta t}_1$ and ${\Delta\phi}_2 = {\Delta\phi}_1$.  Because ${\Delta t}_2 = {\Delta t}_1$, the time between two successive maxima of the mass flow rate at the stellar surface is the same as at the clump, where it is $\DtB$, and hence the frequency $\nu_1$ of the luminosity oscillation is $1/\DtB = \nuorb-\nus \equiv \nuB$ (see eq.~[\ref{nu1}]). Because the rate at which the azimuthal phase of the footprint advances is the same as the rate at which the azimuthal phase of the clump advances, the rotation frequency $\nu_2$ of the footprint is $(\phi_2-\phi_1)/(t_2-t_1) = \nuorb$ (see eq.~[\ref{nu2}]) and the difference between $\nu_2$ and $\nu_1$ is therefore $\nus$.

Suppose now that the inward motion of the clumps near $\rsp$ is too large to be neglected. The radial distance of a given clump decreases with time and therefore the time required for gas to spiral inward from the clump to the stellar surface steadily decreases. Hence ${\Delta t}_2 < {\Delta t}_1$. Equation~(\ref{nu1}) shows that to first order in $\vcl$, the frequency $\nu_1$ of the lower kilohertz QPO is $\nuB(1+\vcl\partial_r\Delta t) > \nuB$; $\nu_1$ is greater than $\nuB$ because the inward motion of the clump reduces the spatial separation between successive maxima of the mass flow rate, analogous to the Doppler shift of a sound wave. Equation~(\ref{nu2}) shows that to first order in $\vcl$, the frequency $\nu_2$ of the upper kilohertz QPO is $\nuorb(1+\vcl\partial_r\Delta t) - \vcl\partial_r\Delta\phi - \nuB^{-1}\vcl\partial_r\nuorb$. Numerical simulations of the gas flow and radiation transport in full general relativity similar to those presented in MLP98 show that the term proportional to $\partial_r\Delta\phi$ is generally the dominant correction term. This term describes the effect on $\nu_2$ of the fact that the gas falling inward from the clump winds a progressively smaller distance around the star as the clump drifts inward, because the gas falls through a smaller radial distance before colliding with the stellar surface. Hence $\nu_2$ is generally less than $\nuorb$. The fractional decrease in $\nu_2$ is typically $\sim50$\% of the fractional increase in $\nu_1$. The increase in $\nu_1$ and the decrease in $\nu_2$ reduce the frequency separation $\Dnu$, which is therefore less than $\nus$.

{\em Inward motion of clumps and gas}.---The previous analysis shows that $\Dnu$ is less than $\nus$. A model of the inward drift of clumps near $\rsp$ and of the infall of gas from the clumps is required in order to compute the detailed behavior of $\nu_1$ and $\nu_2$. Here we consider a simple model that captures the basic physics involved.

Let $\vcl$ be the inward radial velocity of clumps near $\rsp$, let $v_g$ be the characteristic initial inward radial velocity of the gas stripped from a clump, and assume that $\vcl \ll v_g$. Then $\nu_1 \approx \nuB/(1-\vcl/v_g)$ (see eq.~[\ref{nu1}]), because the motion near the clump contributes most to $\Delta t$. Based on the results of the numerical simulations cited above, we assume that the fractional decrease in $\nu_2$ is equal to one-half the fractional increase in $\nu_1$. Then $\nu_2 \approx \nuorb (1-{\scriptstyle{1\over 2}}\vcl/v_g)$. Our simulations indicate that the initial inward radial velocity of stripped gas is fairly insensitive to the distance of the clump from the star and hence for simplicity we assume that $v_g$ is independent of $\rsp$. We also assume for simplicity that $\vcl$ and $v_g$ are approximately constant during the lifetime of a clump.

The inward radial velocity of a clump with angular momentum $J$ caused by removal of its angular momentum by a braking torque $N$ is $\vcl=N/(\partial J/\partial r)$. For a clump of mass $m$ in a nearly circular orbit in the Schwarzschild spacetime,
${\partial J/\partial r} = (m/2)(r-6M)(r-3M)^{-3/2}$ and hence 
 \begin{equation} 
 \label{vcl}
 \vcl(\rsp) =
 \frac{(2/m)
 \left(\rsp-3M\right)^{3/2}}{\left(\rsp-6M\right)}N
 \;, 
 \end{equation}
in units with $G=c=1$. This expression for $\vcl$ diverges at $\rsp=6M$ because the inertia of the clump has been neglected. In reality, $\vcl(\rsp)$ increases as $\rsp$ approaches the radius of the ISCO but remains finite (see Fig.~\ref{fig:vr}). 

There are several effects that may produce a braking torque $N$ on orbiting clumps. Here we consider one likely possibility. $\rsp$ lies inside the star's magnetosphere (MLP98), where the pressure of the stellar magnetic field compresses gas clumps, isolating them from one another. We therefore assume that clumps near $\rsp$ are in pressure equilibrium with the stellar magnetic field, that they interact only weakly with each other, and that $N$ is dominated by the interaction of the clump with the stellar magnetic field (see Ghosh\& Lamb 1979; Aly \& Kuipers 1990; Ghosh \& Lamb 1991).\break

 \vbox{\vskip -0.2truecm 
 \centerline{\epsfxsize=7truecm
 \epsfbox{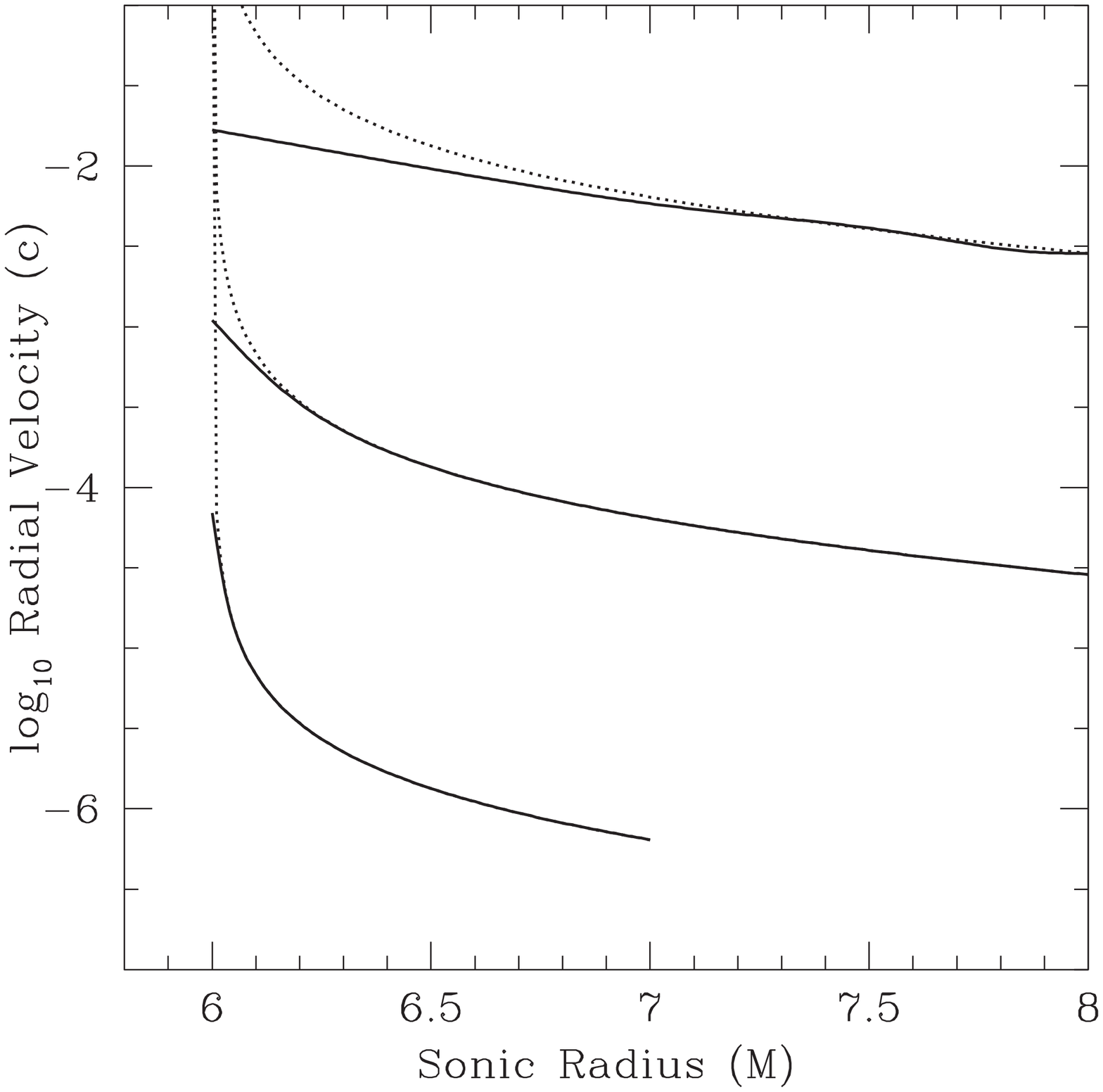}}
 \vskip -0.0truecm
 \figcaption[]
 {Inward radial velocity $\vcl$ of clumps orbiting near the sonic radius $\rsp$ as a function of $\rsp$, computed numerically (solid lines) and using equation~(\ref{vcl}) (dotted lines), for (top to bottom) $c_N=10^{-3}$, $10^{-5}$, and $10^{-7}$ in units of the star's gravitational mass $M$.\label{fig:vr}}
}
 \vskip 0.4truecm

\noindent     Then $N = \gamma \pi a^2\rsp B_{\rm sp}^2$, where $\gamma$ is a dimensionless factor where $\gamma$ is a dimensionless factor $\sim 1$ that depends on the distortion of the stellar magnetic field by the clump and the clump's shape, $a$ is the mean radius of the clump, and $B_{\rm sp}$ is the magnetic field at $\rsp$. Radiation from the star is likely to maintain clumps at the local Compton temperature, which is almost independent of the distance to the star; we therefore assume that the clump temperature is approximately independent of $\rsp$. For simplicity, we also assume that $m$ is approximately independent of $\rsp$. Then $(3/4\pi)2m k_BT/m_p a^3 = B^2/8\pi = \mu^2 \rsp^{-6}$ for a dipolar field with moment $\mu$, where $m_p$ is the mass of a proton, and hence $a \propto \rsp^2$ and $N \propto \rsp^{-1}$. With these assumptions, the evolution equation for the specific angular momentum $u_\phi$ of a clump is $du_\phi/d\tau = c_N/r$, where the torque coefficient $c_N$ (which like $u_\phi$ and the proper time of the clump has the dimensions of a mass) depends on geometrical factors, fundamental constants, and $\gamma$. The evolution equation for the radial position of a clump moving in the Schwarzschild spacetime in the absence of significant nongravitational radial forces is
 \begin{equation}
 {d^2r\over{d\tau}}=-{M\over r^2}+\left(1-{3M\over r}\right)
 {u_\phi^2\over r^3} \;.
 \end{equation}
We computed $\vcl(\rsp)$ by integrating numerically the clump equations of motion, starting with the values of ${\dot r}$ and ${\ddot r}$ given by equation~(\ref{vcl}) at large $\rsp$. 

Figure~\ref{fig:vr} shows how $\vcl$ varies with $\rsp$ for $c_N=10^{-7}$, $10^{-5}$, and $10^{-3}$ in units of the star's gravitational mass $M$. These $c_N$ values correspond to clump densities $\sim1$--10 times the mean density in the disk, for $B_{\rm sp} \sim 10^8$--$10^9\,$G. The inward velocity of clumps increases as $\rsp$ approaches the ISCO at $6M$, but remains finite. For plausible values of $c_N$, $\vcl$ is small enough even near the ISCO that the inspiral time of the clumps there is long enough to be consistent with the observed coherence of the kilohertz QPOs.

\section{RESULTS AND DISCUSSION}

The model developed here of the effect of the inward drift of clumps on the frequencies of the kilohertz QPOs has three parameters: the gravitational mass $M$ and spin frequency $\nus$\break

 \vbox{ \vskip -0.30truecm
 \centerline{\epsfxsize=9truecm
 \epsfbox{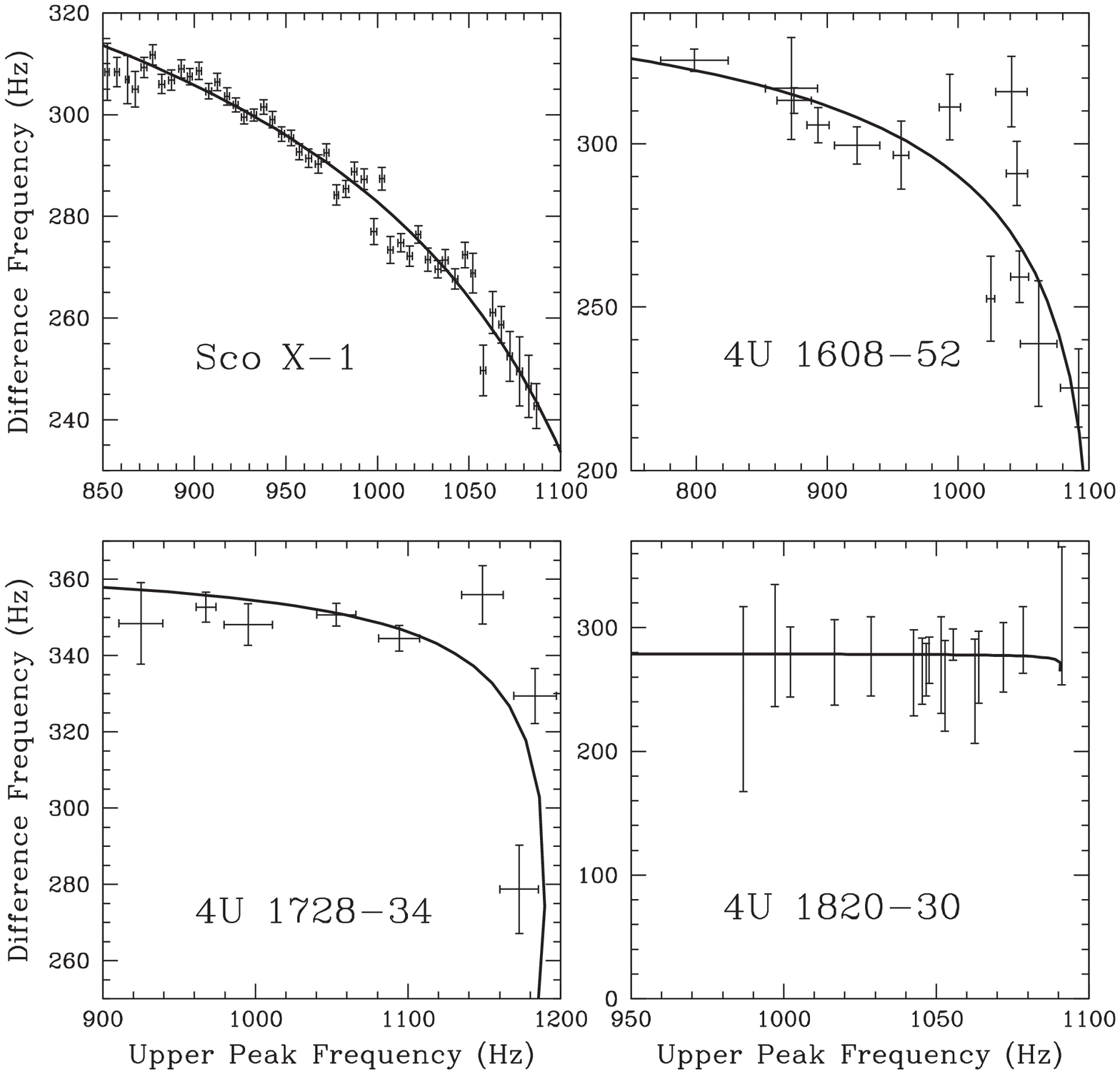}}
 \vskip -0.15truecm
 \figcaption[]{Comparison of fits of the SPBF gas dynamical model described in the text (solid curves) with the $\Delta\nu$--$\nu_2$ correlations observed in four sources. The best-fitting parameter values are \hbox{Sco~X-1}: $M=1.59\,\msun$, $\nus=352\,$Hz, $c_N=5\ee{-3}$, $\chisqdof=95.5/46$; \hbox{4U~1608$-$52}: $M=1.76\,\msun$, $\nus=343\,$Hz, $c_N=2.8\ee{-3}$, $\chisqdof=16.9/10$; \hbox{4U~1728$-$34}: $M=1.745\,\msun$, $c_N=6\ee{-4}$, $\chisqdof=8.5/6$; and \hbox{4U~1820$-$30}: $\nus=279\,$Hz, $c_N = 10^{-5}$, $\chisqdof=7.0/18$. For \hbox{4U~1728$-$34}, $\nus$ was set equal to the 364~Hz frequency of its burst oscillations; for \hbox{4U~1820$-$30}, $M$ was set equal to $2.0\,\msun$, the value indicated by the plateau in its frequency-flux relation (see text).\label{fig:Dnu}}
 }
 \vskip 0.4truecm

\noindent       of the neutron star and the coupling coefficient $c_N$. If $M$ or $\nus$ are known, there are only two parameters. Figure~2 shows the best fits of the model to the frequencies of the kilohertz QPOs in \hbox{Sco~X-1}, \hbox{4U~1608$-$52}, \hbox{4U~1728$-$34}, and \hbox{4U~1820$-$30}. The model gives good fits ($\chisqdof$ of 0.4 to 2.1, not including systematic errors), masses (1.59 to $2.0\,\msun$), spin rates (279 to 364$\,$Hz), and coupling coefficients for all four sources.

The refined version of the SPBF model presented here is consistent with our previous prediction (MLP98) of a plateau in the kilohertz QPO frequency--X-ray luminosity relation when the sonic radius approaches the ISCO. In particular, the fit to the \hbox{4U~1820$-$30} data shown in Figure~\ref{fig:Dnu} is consistent with the interpretation that the plateau in the QPO frequency vs.\ X-ray flux relation reported in this source (Zhang et al.\ 1998; Kaaret et al.\ 1999) indicates that (1)~there is an ISCO around this neutron star and (2)~the  asymptotic frequency of the upper kilohertz QPO is the frequency of the ISCO.

Finally, we note that the fundamental frequency in the SPBF model is the orbital frequency at the sonic radius, which is typically $\sim1,000\,$Hz. Hence, even the largest observed difference between $\Dnu$ and $\nus$ ($70\,$Hz in \hbox{4U~1728$-$34}) and the largest observed variation of $\Dnu$ ($100\,$Hz in \hbox{4U~1608$-$52}), although highly significant, represent only very small (5\%--7\%) deviations of the kilohertz QPO frequencies from the values computed in the circular orbit approximation used previously.

In summary, we have shown that in the sonic-point beat-frequency model, the inward radial motion of the accretion flow causes the frequency separation $\Dnu$ between the kilohertz QPOs to be less than the spin frequency $\nus$ of the neutron star. $\Dnu$ is equal to $\nus$ only if the inflow velocity at the sonic radius is negligible. The inflow velocities expected at $\rsp$ naturally produce the small ($\sim$0.2\%--5\%) corrections to the frequencies predicted for circular orbits (MLP98) needed to explain the differences between $\Dnu$ and $\nus$ and the decreases of $\Dnu$ with increasing $\nu_2$ observed in some sources. We have developed a simple physical model of the effect of inflow on the frequencies of the upper and lower kilohertz QPOs. The model fits well the frequencies of the kilohertz QPOs in four Z and atoll sources where they have been fairly accurately measured and gives reasonable values for the masses and spin rates of the neutron stars in these sources. 

We emphasize that the frequencies of the kilohertz QPOs predicted by the SPBF model depend sensitively on the properties of the spacetime near the neutron star, including the gravitomagnetic torque produced by the spinning neutron star and whether there is an innermost stable circular orbit. Hence, if the sonic-point model is shown to be correct, measurements of the frequencies of the kilohertz QPOs can be used not only to determine the properties of neutron stars but also to detect and measure gravitational effects in the strong-field regime (see MLP98).

\acknowledgements

We thank Michiel van der Klis, Mariano M\'endez, and Will Zhang for extensive discussions. We gratefully acknowledge the hospitality of the Aspen Institute for Physics, where many of these discussions took place. We are especially grateful to Mariano M\'endez for providing the \hbox{Sco~X-1}, \hbox{4U~1608$-$34}, and \hbox{4U~1728$-$34} data and to Will Zhang for providing the \hbox{4U~1820$-$30} data. This research was supported in part by NSF grant AST~96-18524 and NASA grant NAG~5-8424 at Illinois and NASA grant NAG~5-9756 at Maryland.


\begin{references}
\def\ref{\par\noindent\hangindent 15pt}

\ref Alpar, A., \& Shaham, J. 1985, Nature, 316, 239

\ref Aly, J.-J., \& Kuipers, J. 1990, A\&A, 227, 473

\ref Ford, E.\,C., van der Klis, M., van Paradijs, J., M\'endez, M., Wijnands, R.\,A.\,D., \& Kaaret, P. 1998, ApJ, 508, L155

\ref Ghosh, P., \& Lamb, F.\,K. 1979, ApJ, 232, 259

\ref Ghosh, P., \& Lamb, F.\,K. 1991, in Neutron Stars: Theory and Observation, ed. J. Ventura \& D. Pines (Dordrecht: Kluwer), 363

\ref Kaaret, P., Piraino, S., Bloser, P.\,F., Ford, E.\,C., Grindlay, J.\,E., Santangelo, A., Smale, A.\,P., \& Zhang, W. 1999, ApJ, 520, L37

\ref Lamb, F.\,K., Miller, M.\,C., \& Psaltis, D. 1998, in The Active X-ray Sky, ed. L.~Scarsi, H.~Bradt, P.~Giommi, and F.~Fiore (Amsterdam: Elsevier), 113 (astro-ph/9802089)

\ref Lamb, F.\,K., Shibazaki, N., Alpar, A., \& Shaham, J. 1985, Nature, 317, 681

\ref Lorimer, D.\,R. 2000, to appear in The Neutron Star -- Black Hole Connection, NATO ASI meeting, Elounda, Crete, 7--18 June 1999 (astro-ph/9911519)

\ref M\'endez, M., \& van der Klis, M. 1999, ApJ, 517, L71

\ref M\'endez, M., \& van der Klis, M. 2000, MNRAS, in press (astro-ph/0006243)

\ref M\'endez, M., van der Klis, M., \& van Paradijs, J. 1998, ApJ, 506, L117

\ref M\'endez, M., van der Klis, M., Wijnands, R.\,A.\,D., Ford, E.\,C., van Paradijs, J., \& Vaughan, B.\,A. 1998, ApJ, 505, L23

\ref Miller, M.\,C. 1999, ApJ, 515, L77

\ref Miller, M.\,C. 2000, submitted to Astr. Lett. Comm. (astro-ph/0007287)

\ref Miller, M.\,C., Lamb, F.\,K., \& Cook, G.\,B. 1998, ApJ, 509, 793

\ref Miller, M.\,C., Lamb, F.\,K., \& Psaltis, D. 1998, ApJ, 508, 791 (MLP98)

\ref Psaltis, D. et al. 1998, ApJ, 501, L95

\ref Strohmayer, T.\,E., \& Markwardt, C.\,B. 1999, ApJ, 516, L81

\ref van der Klis, M. 2000, ARA\&A, in press (astro-ph/0001167)

\ref van der Klis, M., Wijnands, R.\,A.\,D., Horne, K., \& Chen, W. 1997, ApJ, 481, L97

\ref Zhang, W., Smale, A.\,P., Strohmayer, T.\,E., \& Swank, J.\,H. 1998, ApJ, 500, L171

\end{references}
\end{document}